\documentclass[]{JFM-FLM_Au}

\usepackage{graphicx}

\usepackage{hyperref}
\hypersetup{
    colorlinks = true,
    urlcolor   = blue,
    citecolor  = black,
}

\usepackage{overpic}[grid=True]

\usepackage{siunitx}
\usepackage[dvipsnames]{xcolor}

\usepackage[english]{babel}
\usepackage[autostyle, english = american]{csquotes}
\MakeOuterQuote{"}

\newcommand{\new}[1]{{#1}} 
\newcommand{\neww}[1]{{#1}} 

\lefttitle{Ruth and Coletti}
\righttitle{ }

\title{Free-surface curvature and its relation to subsurface turbulence}

\author{Daniel J. Ruth\aff{1} \and Filippo Coletti \aff{1}}

\affiliation{\aff{1}Institute of Fluid Dynamics, ETH Zurich, Zurich, Switzerland}

\corresau{Daniel J. Ruth, daruth@ethz.ch}

\usepackage{fancyhdr}

\fancypagestyle{arxivfirst}{
    \fancyhf{} 
    \chead{\small \textit{Under consideration for publication in J. Fluid Mech.}}
    \cfoot{\thepage} 
}

\fancypagestyle{arxivrest}{
    \fancyhf{}

    \cfoot{\thepage}
}

\begin{document}
\maketitle
\thispagestyle{arxivfirst}

\begin{abstract}
\new{The free surface atop a turbulent liquid flow is deformed by the underlying fluid motion, with the turbulence imprinting its geometry on the surface. Here we develop a theoretical framework to model such deformations based on the Euler equation with gravity and surface tension, and evaluate it against simultaneous high-resolution measurements of surface topography and subsurface velocity fields in a zero-mean-flow turbulent water tank. We consider a range of Reynolds and Froude numbers, focusing on regimes in which the surface is unbroken. Over a range of spatial and temporal scales, we find close  quantitative agreement between the measured surface curvature and that which is modeled based on the velocity field a few millimeters beneath the surface, both from the statistical and the local/instantaneous standpoints. Importantly, we verify a strong correlation between the magnitudes of the surface curvature and the divergence of the near-surface horizontal velocity, which in turn is directly related to gas and heat transfer at the air-water interface. We discuss how the sub-surface motion at increasing depths decorrelates from the surface shape, and does so more rapidly at smaller spatial scales. These findings demonstrate that measurements of surface deformations may be used to sense the state of the flow beneath the surface and provide a foundation to make optically-based inferences of processes controlled by near-surface turbulence.}
\end{abstract}

\begin{keywords}
Authors should not enter keywords on the manuscript, as these must be chosen by the author during the online submission process and will then be added during the typesetting process (see \href{https://www.cambridge.org/core/services/aop-file-manager/file/61436b61ff7f3cfab749ce3a/JFM-Keywords-Sept-2021.pdf.}{Keyword PDF} for the full list).  Other classifications will be added at the same time.
\end{keywords}


\section{Introduction}
\label{sec:introduction}

A free surface heavily impacts the turbulent flow in its vicinity, affecting its velocity fluctuations, velocity gradients, and interfacial energy transfer \citep{hunt_free-stream_1978,magnaudet_high-reynolds-number_2003,guo_interaction_2010,ruth_structure_2024,wu_localised_2026}. In the limit of gravity and/or surface tension fully outweighing the subsurface pressure fluctuations, the surface remains virtually flat, while it can deform and even break for sufficiently large spatial and velocity scales of the turbulence: pressure variations in the liquid locally raise and lower the water surface overhead, yielding topographical imprints that can be related to distinct flow features beneath \citep{longuet-higgins_surface_1996,brocchini_dynamics_2001}. For example, circular depressions (dimples) clearly reflect the decreased pressure at the core of surface-attached vortices \citep{babiker_vortex_2023}. 

Such a connection between free surface deformation and properties of the turbulent flow beneath remained largely speculative until experimental and numerical advances in the last two decades enabled systematic investigations. \cite{savelsberg_experiments_2009} simultaneously measured the surface deformations and the flow in an open channel where turbulence was forced by an active grid. They investigated correlations between the surface slope and quantities derived from subsurface velocity gradients, and found them to be substantially weaker than in the simpler case of vortices shed behind a cylinder. \cite{guo_interaction_2010} studied the free surface above homogeneous isotropic turbulence via direct numerical simulations, and found that its motion is characterized by both propagating waves and turbulence-generated deformations. The same numerical configuration was used by \cite{babiker_vortex_2023} and \cite{aarnes_vortex_2025}, demonstrating the link between surface features (such as dimples and scars) and near-surface vortices. \cite{babiker_experimental_2026} used a jet-stirred water tank to show the connection between the surface topography and an integrated measure of the divergence of the surface-parallel velocity field below the surface. In the limit of vanishingly small depth (and weak slope of the surface deformation) such divergence becomes the surface divergence – a critical quantity in models of turbulence-enabled interfacial gas transfer, whose value dictates the extent to which turbulence replenishes the surface with water from beneath \citep{mckenna_role_2004,turney_airwater_2013,herlina_direct_2014}. Simulations of strong free-surface turbulence by \cite{calado_interfacial_2025} showed visual correlation between vorticity magnitude and surface curvature. Recent experiments by \cite{berhanu_free-surface_2026} and simulations by \cite{foggi_rota_effect_2026} found that the elevation spectrum of the surface above turbulent water is consistent with that of canonical turbulent pressure fluctuations modulated by surface tension and gravity. The abovementioned studies considered approximately homogenous subsurface turbulence, which is also the focus of the present paper. Several recent works have also studied the surface manifestation of underwater disturbances such as cylinder arrays \citep{mandel_surface_2019} and steps \citep{luo_water_2023}.

The above indicates that characterizing the free surface topography is a promising avenue for sensing the state of the subsurface flow – a lofty goal in field settings. Experimental techniques to reconstruct surface elevation maps with high resolution and accuracy, suitable for laboratory measurements, have been devised and improved in recent years \citep{cobelli_global_2009,moisy_synthetic_2009,gakhar_extracting_2022,gomit_free-surface_2022,bheeroo_comparison_2023,semati_simultaneous_2026}. Also, recent advances in observational approaches and data-driven algorithms suggest exciting progress in this area. For instance, high-resolution measurement of small-scale surface deformations on open bodies of water are enabled by polarimetric slope sensing \citep{laxague_e-pss_nodate}. At the same time, data-driven methods such as the one proposed by \cite{gakhar_extracting_2022}, \cite{xuan_reconstruction_2023}, and \cite{moen_mapping_2025} based on neural networks can enable and accelerate the reconstruction of the near-surface flow from the surface elevation and/or velocity. In this scenario, a framework founded on first principles to connect surface topography and subsurface flow is most desirable.

In the present study we seek further understanding of the relationship between surface deformations and subsurface turbulence, both from statistical and local/instantaneous standpoints. Leveraging the large-scale laboratory facility introduced in \cite{ruth_structure_2024}, we apply time-resolved particle image velocimetry (PIV) and background-oriented schlieren (BOS) to obtain simultaneous deformation and velocity fields at high spatial and temporal resolution, needed to resolve the wide range of scales associated to the homogeneous turbulence generated at depth. The surface curvature emerges as a natural quantity to connect the surface and subsurface dynamics, particularly the horizontal divergence of the velocity field. The rest of the paper is structured as follows. Section \ref{sec:theoretical_framework} details the inviscid approach by which we relate near-surface velocity gradients to surface curvature, accounting for effects of both gravity and surface tension. Section \ref{sec:methods} details the experimental methods, describing the apparatus, the considered regimes, and the measurement approach. A statistical description of the surface topology and subsurface flow, and the relation between their time-averaged statistical values, is given in sections \ref{sec:structure_statistical} - \ref{sec:variance_vs_variance}. In sections \ref{sec:modeling_gravcap} - \ref{sec:depth_dependence}, we investigate the local and instantaneous relationship between the surface and subsurface: we first correlate the surface curvature with the terms contributing to the pressure disturbances just beneath the surface, and then we investigate how the surface-subsurface correlation is affected by the resolved spatial scales and depth. We discuss our findings in section \ref{sec:discussion_conclusions}, concluding with an outlook on the prospect of applying these results to remote sensing techniques.

\section{Theoretical framework}
\label{sec:theoretical_framework}

We start from the inviscid Navier-Stokes (Euler) equation evaluated at the free surface. As detailed below, we relate the surface curvature $\kappa = -\nabla^2 \eta$ (with $\eta$ the surface elevation) to the subsurface flow through inviscid dynamics: pressure fluctuations conforming to the velocity field are related to the surface elevation through the effects of gravity and capillarity. \cite{savelsberg_experiments_2009} attempted to relate the surface slope $\vec{\nabla}\eta$ to the subsurface flow in a similar inviscid manner. Here the time-resolved measurements reported allow us to include unsteady effects which have been neglected in previous analyses. 

The Euler equation, differentiated spatially after including the Laplace pressure term, yields the following partial differential equation for the surface curvature:
\begin{equation}
    g \kappa - \frac{\sigma}{\rho} \nabla^2 \kappa = - \frac{D \beta_0}{D t} - \beta_0^2 + 2 q_0 \label{eq:kappa_invariantdecomp},
\end{equation}
which we will refer to as the Euler-Laplace equation. Here $g$ is the gravitational acceleration in direction $-z$, $\sigma$ is the surface tension, and $\rho$ is the water density. The right-hand-side terms involve the invariants of the velocity gradient tensor $A_{ij} = \partial u_i / \partial x_j$ (with indices $i$ and $j$ rotating through the surface-parallel directions $x$ and $y$), $\beta = -\partial u_z / \partial z$  is the surface-parallel divergence, and $q=\det(A_{ij})$. The subscript 0 denotes quantities measured at the free surface ($z=\eta$, with $\eta=0$ when the surface is undeformed). The origin of the Cartesian coordinates is at the undisturbed water surface. The fluid velocity has components $u_x$, $u_y$ and $u_z$ in the surface-parallel and vertical directions, respectively. In using the letter $\beta$ we follow the bulk of the literature on free-surface turbulence; we will also use $p=-\beta$, in keeping with the classic notation describing the velocity gradient tensor \citep{perry_description_1987,cardesa_invariants_2013}. Here and in the following, the nabla operator is applied only in the horizontal direction. Over the considered range of relatively small Froude and Weber number (defined in section \ref{sec:flow_regimes}), the surface deformation is sufficiently small that derivatives in the horizontal and surface-parallel directions are quantitatively close \citep{guo_interaction_2010,calado_interfacial_2025}. We therefore use the horizontal derivative such that the analysis can be extended to arbitrary depths.

To investigate the relationship between surface deformation and the subsurface flow, we use equation \ref{eq:kappa_invariantdecomp} to define $\kappa_z$ as the surface curvature modeled from the pressure field conforming to the velocity at a depth $z$. This yields
\begin{equation}
    \mathcal{L} \left[ \kappa_z \right] = - \frac{D \beta}{D t} - \beta^2 + 2 q \label{eq:kappaz_transformed_invariantdecomp}
\end{equation}
where, for the generic quantity $a$ evaluated at depth $z$, we use the operators $\mathcal{L} [a] = g a + (\sigma/\rho) \nabla^2 a $ and $D a / D t= \partial a/\partial t + \vec{u}\cdot \vec{\nabla} a$. The modeled curvature is then solved for as
\begin{equation}
    \kappa_z = - \widehat{\frac{D \beta}{D t}} - \widehat{\beta^2} + 2 \widehat{q} \label{eq:kappaz_invariantdecomp}
\end{equation}
where $\hat{a} = \mathcal{L}^{-1}[a]$. Naturally, as the depth increases, the velocity field along the horizontal plane decorrelates from the surface velocity field, and the accuracy of equation \ref{eq:kappaz_invariantdecomp} is expected to worsen. This will be discussed in section \ref{sec:depth_dependence}. 

The above parameterization in terms of $\beta$ and $q$, two invariants of the surface-parallel velocity gradient tensor \citep{perry_description_1987}, provides a direct link between the surface topography and the underlying flow topology. A universal shape for the joint probability distribution function (JPDF) of $p=-\beta$ and $q$ has been found for planar cuts through homogeneous three-dimensional (3D) turbulent flows, with a teapot shape highlighting the predominance of vortex stretching \citep{cardesa_invariants_2013}. In the vicinity of a minimally deformed free surface, \cite{qi_restricted_2025,qi_small-scale_2025,wu_localised_2026} showed how this JPDF becomes more symmetric, signaling the balance of vortex stretching and compression due to downwellings and upwellings, respectively. 

The right-hand side of equation \ref{eq:kappaz_invariantdecomp} can be reformulated in terms of gradient-based flow quantities canonically associated to surface flow features. Namely, we define the vertical (surface-normal) vorticity $\omega_z = \partial u_x / \partial y - \partial u_y / \partial x$ and the squared strain rate $s^2 = (\partial u_x / x - \partial u_y/y)^2 + (\partial u_x/\partial y + \partial u_y / \partial x)^2$. Expressed in terms of these variables, the modeled curvature is
\begin{equation}
    \kappa_z = - \widehat{\frac{D \beta}{D t}} - \frac{\widehat{\beta^2}}{2} + \frac{\widehat{\omega_z^2}}{2} - \frac{\widehat{s^2}}{2}. \label{eq:kappaz}
\end{equation}

To gain immediate physical insight in the relationship among the quantities, let us neglect for the moment the effect of surface tension, which in many turbulent flows (including the one we consider in the present study) is small compared to that of gravity. In this limit, each $\hat{\alpha}$ term becomes simply proportional to $\alpha$ through the gravitational acceleration, leading to the relation
\begin{equation}
    \kappa_\mathrm{g} = - \frac{1}{g} \left( \frac{D \beta}{D t} - \frac{\beta^2}{2} + \frac{\omega_z^2}{2} - \frac{s^2}{2} \right) = - \frac{1}{g} \left( \frac{D \beta}{D t} - \beta^2 + 2 q \right) \label{eq:kappa_g}
\end{equation}
where the subscript "g" denotes “gravity-dominated”. That $\kappa_\mathrm{g} \propto - \beta^2/g$ indicates how both upwellings ($\beta>0$) and downwellings ($\beta<0$) yield negative curvature, i.e., a local bulge in the surface due to the increased pressure at the stagnation point associated with both types of events. Similarly, $\kappa_\mathrm{g} \propto \omega_z^2 /g$ is consistent with fluid rotation decreasing local pressure and leading to positive curvature, e.g., dimples sitting atop surface-attached vortices \citep{babiker_vortex_2023}. In this gravity-dominated scenario, we note that the link between the curvature and the forcing terms is independent of the length scale of the turbulence. Considering a stand-alone circular and uniform upwelling as an example, the radial acceleration along the surface creates a pressure gradient which increases linearly away from the center, resulting in a quadratic pressure profile (which yields a quadratic elevation profile and thus a constant surface curvature). 

In the following we will use the Euler-Laplace equation (\ref{eq:kappa_invariantdecomp}) and its various versions (\ref{eq:kappaz_invariantdecomp}-\ref{eq:kappa_g}) as a framework to interpret experimental observations linking surface topography and subsurface flow topology. 

\section{Methods}
\label{sec:methods}

\subsection{Experimental apparatus}
\label{sec:experimental_apparatus}

\begin{figure}
  \centerline{\begin{overpic}[width=1\linewidth]{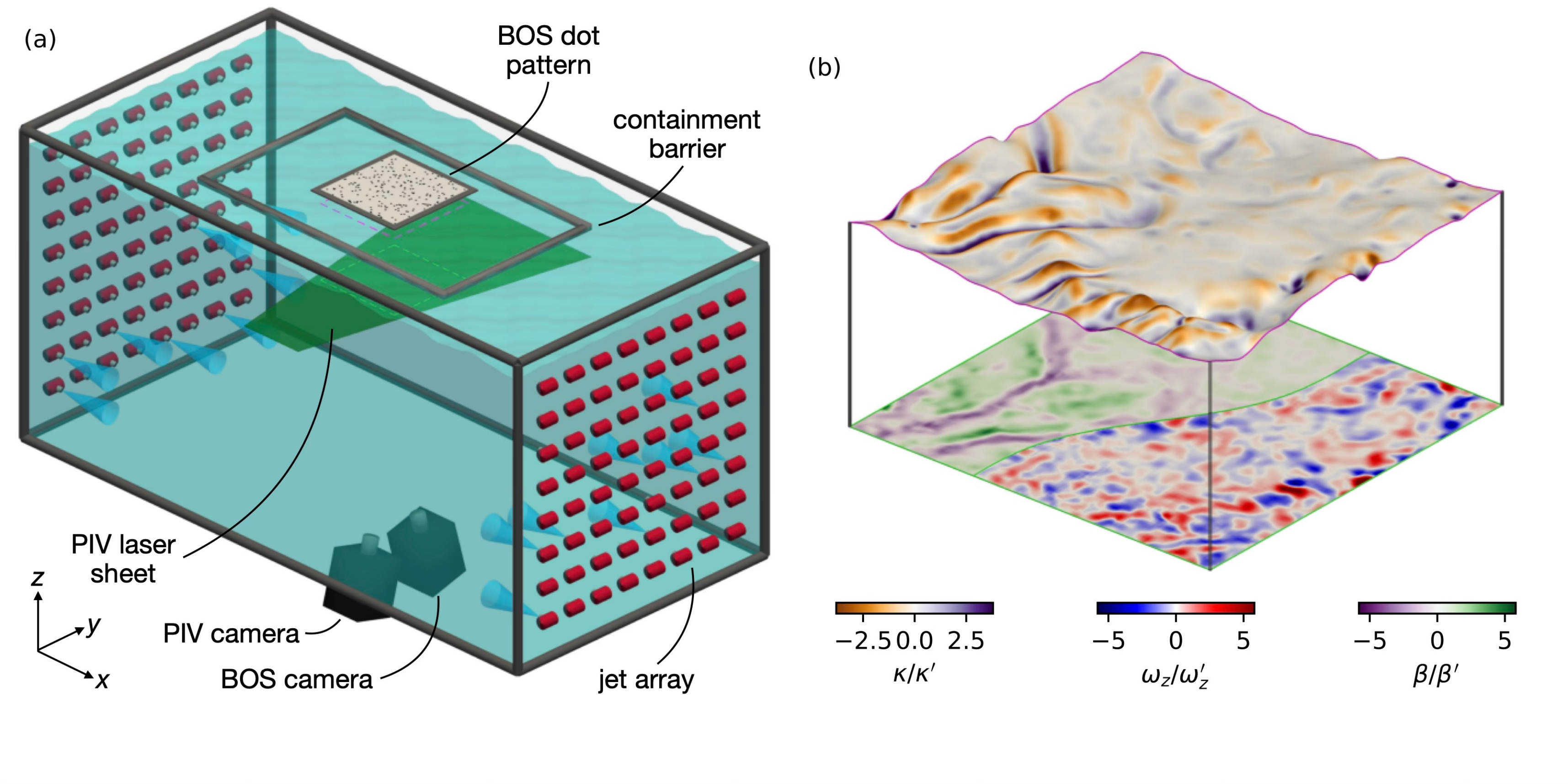}\end{overpic}}
  \caption{The generation and measurement of turbulence and the resulting surface deformations. (a) Schematic of the experiment. Turbulence is generated by turning pumps in two opposing randomly-actuated jet arrays on and off. The flow in a horizontal plane beneath the surface is measured by shining a horizontal light sheet (in green) at some depth and imaging with a PIV camera beneath the tank. Simultaneously, a random dot pattern is held just above the water surface, and a BOS camera beneath the tank images the apparent displacement of the dots due to the water surface deformation. (b) A visualization of the concurrently-resolved surface topography and sub-surface velocity fields. The surface elevation (obtained by integrating the surface gradient) is shown (vertically stretched) and is colored by its curvature. Beneath, the vertical vorticity and horizontal divergence in a plane \SI{2}{mm} beneath the surface are shown.}
\label{fig:schematic}
\end{figure}

We perform experiments in which turbulence is forced beneath an air-water interface and the subsurface flow and surface deformations are measured simultaneously. A schematic of the apparatus, described in detail in \cite{ruth_structure_2024} and \cite{li_relative_2024} and just briefly reintroduced here, is shown in figure \ref{fig:schematic} (a). Turbulence with near-zero mean flow is generated in a \SI{2}{m^3} tank of water using two facing arrays of randomly actuated jets \citep{bellani_homogeneity_2014,carter_generating_2016}, each with eight rows and eight columns of pumps separated by \SI{10}{cm}. The sunbathing algorithm proposed by \cite{variano_random-jet-stirred_2008} is employed, with the pumps turned on and off for mean durations of \SI{3}{s} and \SI{21}{s}, respectively. The present setup is considerably larger than other similar installations used to investigate free-surface turbulence \citep{variano_turbulent_2013,jamin_experimental_2025,babiker_experimental_2026} and allows accessing a wider range of turbulent Reynolds numbers (detailed below). The flow quality (including the bulk level of homogeneity, small magnitude of mean velocity and mean shear, and small-scale isotropy; see \cite{ruth_structure_2024} and \cite{wu_localised_2026}) provides a flow configuration that well approximates the canonical case of zero-mean homogenous turbulence interacting with the free surface above.

To vary the intensity of the turbulence, the power supplied to the pumps is controlled via pulse width modulation. As a second control parameter, we also vary the vertical distance between the surface and the region of the water forced by the jets. \new{By turning off the top $n$ of the eight rows of pumps (with $1\leq n \leq 4$ employed in this work), the turbulence is forced up to a distance only $\SI{5}{cm}+ (n \times \SI{10}{cm})$ beneath the surface.} The spatial decay of the turbulence away from the forcing region \citep{hopfinger_spatially_1976} and the blockage effect by the no-penetration boundary condition \citep{hunt_free-stream_1978,ruth_structure_2024,jamin_experimental_2025} reduce the turbulence intensity approaching the surface. 

The inner region of the water surface is contained by a rectangular barrier, \SI{73}{cm} by \SI{45}{cm}, which extends approximately \SI{1}{cm} beneath the surface. The barrier ensures that surface waves originating near the jet arrays by the firing of the pumps do not propagate into the measurement region. As the barrier perimeter is much larger than the characteristic length scales of the flow and of the investigated region at its center, its influence on the investigated dynamics is deemed negligible. 

The presence of surfactants may significantly affect free-surface turbulence dynamics \citep{shen_effect_2004,mckenna_role_2004,herlina_direct_2014}. To minimize contamination, in every experiment the surface water inside the barrier is skimmed and returned to the bulk for approximately five minutes before each \SI{90}{s} recording. Dedicated tests indicate that on the order of minutes to hours after such cleaning, the surface deformations begin to decrease in magnitude, which is attributed to the absorption of surfactants from the bulk to the surface. 

\subsection{Measurement approach}
\label{eq:measurement_approach}

The subsurface velocity is measured by time-resolved PIV. A continuous \SI{450}{nm} blue laser beam (CNI Laser) is expanded into a horizontal sheet and held at a depth between $z=\SI{-100}{mm}$ and \SI{-2}{mm}. A Phantom VEO 640 camera beneath the tank, outfitted with a \SI{50}{mm} lens, records pairs of tracer particle images separated by 4 or \SI{5}{ms}, depending on the typical speed of the flow. Image pairs are recorded at \SI{100}{Hz}. The velocity components $u_x$ and $u_y$ are obtained through a multi-pass cross-correlation technique using the software PIVLab \citep{thielicke_pivlab_2014}, yielding a spatial resolution of approximately \SI{2}{mm} over a field of view of $\SI{20}{cm} \times \SI{20}{cm}$. Statistics are based on \num{18000} realizations (two recordings of \SI{90}{s} each), which is verified to suffice for convergence of the reported quantities. We note that surface skimming produces a significant increase in the quality of the PIV data close to the surface. Without this procedure, individual and aggregated tracer particles floating on the surface are visible in PIV images focused on a plane a few mm beneath the surface, despite not being directly illuminated by the laser sheet, negatively impacting the measurement accuracy.

Concurrently to the PIV measurements, the surface deformation is reconstructed using BOS \citep{moisy_synthetic_2009,bullee_influence_2024}. To this end, a second Phantom VEO 640 camera, also outfitted with a \SI{50}{mm} lens, is positioned beneath the tank and focused on a translucent random dot pattern held \SI{20}{mm} above the water surface. The dot pattern is illuminated from above with strobed LED lights. The BOS camera, synchronized with the LEDs, records images at \SI{100}{Hz}, approximately \SI{50}{ms} after each second PIV image is obtained (effectively simultaneous to the flow velocity measurement, \new{as the reported correlations do not change significantly by shifting the flow and surface measurements in time by $\pm \SI{0.01}{s}$)}. The apparent displacements of the dots (due to the deformed surface’s refraction of light) are obtained through cross-correlation with an approach similar to that applied to the PIV image pairs. In the present range of small slopes, \new{$|\vec{\nabla}\eta|' \leq \mathcal{O} ( 10^{-2})$}, the displacement field is linearly related to the surface gradient, \new{with the scaling factor defined for the present optical setup (in which a camera beneath a tank images a dot pattern above the water surface) by \cite{weichert_thesis_2024}}. Here and in the following, the prime denotes the root mean square (r.m.s.) of the fluctuations. The gradient field is differentiated spatially via a central difference scheme to obtain the surface curvature.

Measured velocity and surface deformation fields, obtained at similar resolutions, are interpolated onto a common grid to facilitate the analysis of their dynamic connections. Smoothing with a Gaussian kernel of standard deviation $\SI{1}{mm}$ is applied to both fields to reduce the noise associated to spatial differentiation. \new{Velocity fields are further smoothed in time with a Gaussian kernel of standard deviation $\SI{30}{ms}$ before they are employed to model the curvature.}

\new{Figure \ref{fig:schematic} (b) displays a sample snapshot of the surface topography (colored by the surface curvature) and the flow field \SI{2}{mm} beneath it (colored by the vertical vorticity and horizontal divergence of the flow). The distance between the two planes and the magnitude of the variations in elevation are exaggerated for visual clarity. The surface elevation fields employed in our visualizations are obtained by  integrating the surface gradients with a spectral method.}

\subsection{Flow regimes}
\label{sec:flow_regimes}

Here we summarize the investigated flow regimes based on the properties measured at the reference depth $z=\SI{-5}{mm}$ for each turbulence forcing conditions. In most cases we also make measurements at $z=\SI{-2}{mm}$, but this is not possible in the most intensely turbulent cases, as the moving surface will occasionally deflect the laser sheet. As shown below, the characteristic scale of the velocity fluctuations (on which we base our regime definition) differs only minorly between the two planes.

\begin{figure}
  \centerline{\begin{overpic}[width=1\linewidth]{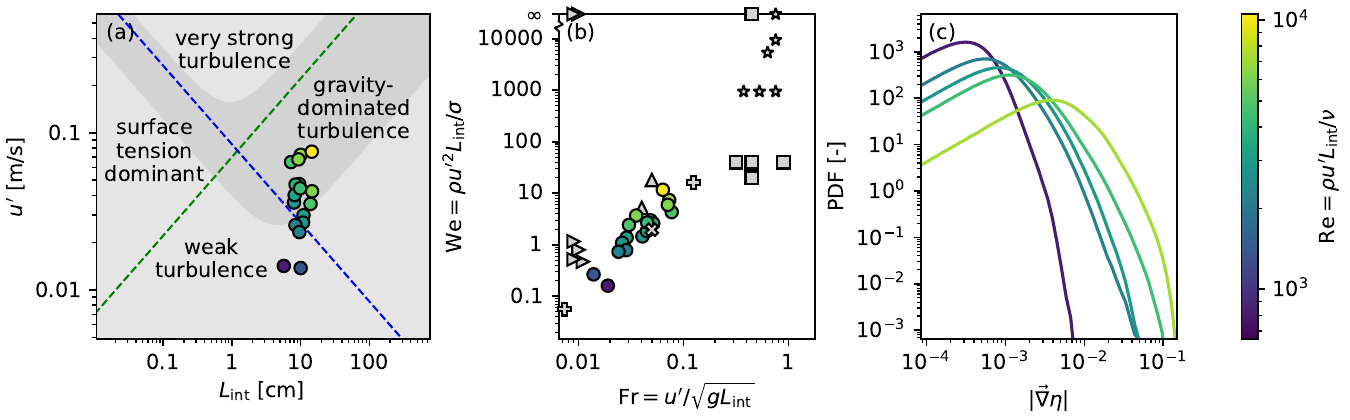}\end{overpic}}
  \caption{Strength of the turbulence generated and the degree of the resulting surface deformations. (a) The large-scale velocity and length scales in a plane \SI{5}{mm} from the surface. (b) The associated Weber, Froude, and Reynolds numbers. Included in panel (b) is dimensionless data from similar experiments and simulations (\cite{guo_interaction_2010}, $\square$; \cite{babiker_experimental_2026}, experimental $\triangle$ and numerical $\triangleright$; \cite{variano_turbulent_2013}, $\times$; \cite{calado_interfacial_2025}, $\star$). (c) For five  cases with varying $\mathrm{Re}$, PDFs of the magnitude of the surface slope.}
\label{fig:parameter_space}
\end{figure}

For each case, characterized by a combination of power supplied to the pumps and number of active jet rows, we compute from PIV the turbulent velocity scale $u' = \sqrt{u_x'^2 + u_y'^2}$. We fit an exponential curve to the longitudinal autocorrelation of velocity fluctuations to obtain the integral length scale $L_\mathrm{int}$ \citep{ruth_structure_2024}. While $L_\mathrm{int}$ remains close to the bulk value of $\sim \SI{10}{cm}$ (which is set by the width of the jets in the centre of the tank), $u'$ spans almost a decade. Based on these two large-scale quantities, figure \ref{fig:parameter_space} (a) positions each investigated case in the parameter space proposed by \cite{brocchini_dynamics_2001}. With weaker forcing we obtain conditions in the “weak turbulence” regime, while stronger forcing pushes us to the “gravity-dominated turbulence” regime. This suggests that, overall, gravity will prevail over surface tension. This is confirmed by the non-dimensional parameter space in figure \ref{fig:parameter_space} (b), which shows the Weber number $\mathrm{We} = \rho u'^2 L_\mathrm{int}/\sigma$ and Froude number $\mathrm{Fr}=u'/\sqrt{g L_\mathrm{int}}$ for each condition, the colour denoting the turbulent Reynolds number $\mathrm{Re}=u' L_\mathrm{int} /\nu$ (where $\nu$ is the water kinematic viscosity). While the Weber number surpasses unity as we transition from weak to gravity-dominated turbulence, the Froude number remains below $0.1$ in all conditions. \new{Corresponding values from relevant studies are given in gray.} Given the relative importance of gravity, we will use the Froude number to characterize the intensity of the near-surface flow. \new{Finally, the probability distribution of the resulting surface slopes are given for five cases with varying turbulence intensities in figure \ref{fig:parameter_space} (c).} Over the range of conditions investigated, \new{the scale of the surface slope fluctuations $|\vec{\nabla}{\eta}|'$ varies between \num{3e-4} and \num{1e-2}.}

\begin{figure}
  \centerline{\begin{overpic}[width=1\linewidth]{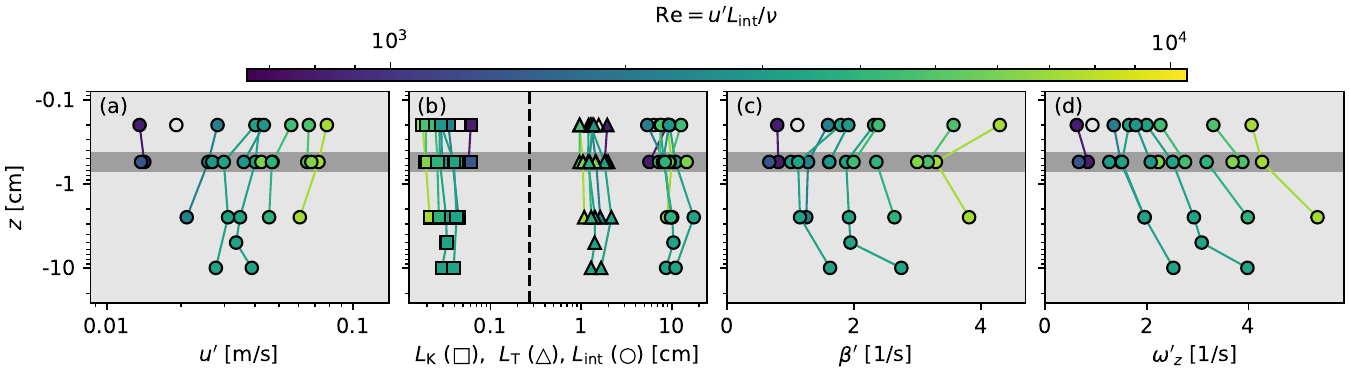}\end{overpic}}
  \caption{Statistical characterization of the turbulence at various depths over all the forcing conditions. (a) rms surface-parallel velocity. (b) Length scales of the turbulence and the gravity-capillary length scale (dashed line). (c) rms horizontal divergence. (d) rms surface-normal vorticity.}
\label{fig:profiles_with_depth}
\end{figure}

Though not the focus of this work, our measurements in various planes indicate how the turbulence varies with depth in the region close to the surface, as shown in figure \ref{fig:profiles_with_depth} where the thick line highlights the reference depth $z=\SI{-5.5}{mm}$. There is a slight increase in the surface-parallel velocity fluctuations near the surface (figure \ref{fig:profiles_with_depth} (a)), an effect attributable to near-surface vertical fluctuations being reoriented to the horizontal \citep{hunt_free-stream_1978}. We evaluate the turbulent energy dissipation rate $\epsilon$ through a fit to the second-order structure function (measured by PIV along the relevant plane), according to \cite{kolmogorov_local_1941}'s theory. We then calculate the Kolmogorov length scale $L_\mathrm{K}=(\nu^3/\epsilon)^{1/4}$ and the Taylor length scale $L_\mathrm{T}=\sqrt{15 \nu u'^2/\epsilon}$. At all conditions and depths, the gravity-capillary scale $L_\mathrm{gc}=\sqrt{\sigma/(\rho g)}=\SI{2.7}{mm}$ sits between the Kolmogorov and Taylor scales (figure \ref{fig:profiles_with_depth} (b)). The horizontal divergence $\beta$ (figure \ref{fig:profiles_with_depth} (c)) first decreases as the turbulence decays away from the forcing region, and then increases approaching the surface, in agreement with numerical simulations \citep{guo_interaction_2010}. The surface-normal vorticity $\omega_z$ (figure \ref{fig:profiles_with_depth} (d)) decreases with $z$ due to the spatial decay of the turbulence from the forcing region.

\section{Structure of the near-surface flow and surface deformation}
\label{sec:structure}

\subsection{Statistical description}
\label{sec:structure_statistical}

\begin{figure}
  \centerline{\begin{overpic}[width=1\linewidth]{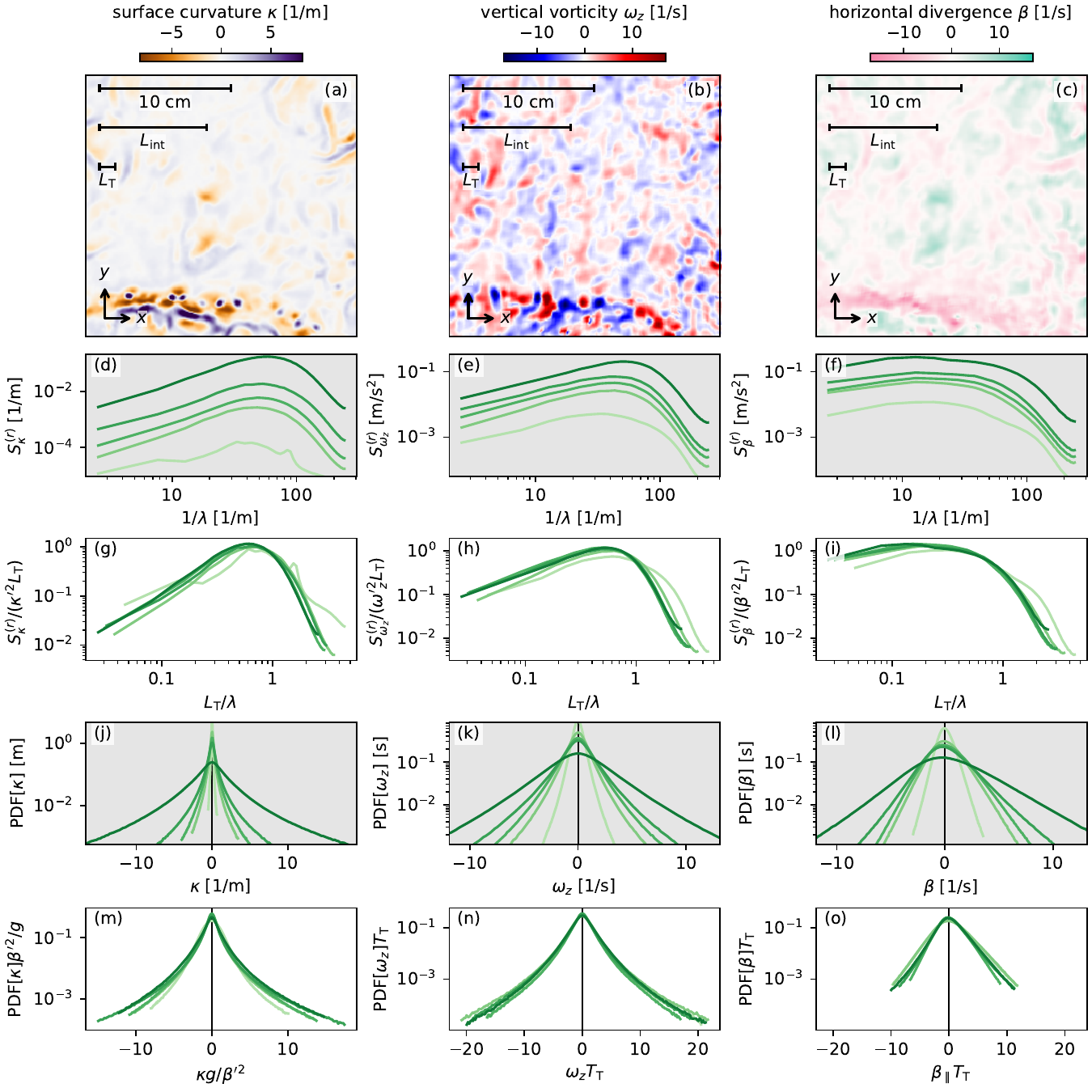}\end{overpic}}
  \caption{Snapshots and statistics of the surface deformations and subsurface flow. (a-c) For one instant in time, snapshots of the surface curvature, and surface-normal vorticity and horizontal divergence 2 mm beneath the surface. Spatial spectra of (d) surface curvature, (e) vorticity, and (f) divergence are shown for five cases with varying Froude number (from light green to dark green, $\mathrm{Fr}=0.019$, 0.028, 0.041, 0.049, and 0.073), and these are nondimensionalized in (g-i). Distributions of the same quantities are shown in panels (j-l), and they are normalized in (m-o).}
\label{fig:spatial_spectra}
\end{figure}

In this section we describe the scale and structure of the surface curvature and how they relate to the statistics of the turbulence beneath. Figures \ref{fig:spatial_spectra} shows a sample snapshot of the surface curvature (a), vertical vorticity (b) and horizontal divergence (c) measured below the surface. Here and in the remainder of this section, we refer to the subsurface plane at $z = \SI{-2}{mm}$, which is roughly at the edge of the viscous layer (whose depth scales as $\mathcal{O}(L_\mathrm{int} \mathrm{Re}^{-1/2})$, \citep{hunt_free-stream_1978}). The integral and Taylor length scales are indicated in each panel. This instantaneous realization illustrates various physical couplings discussed in section \ref{sec:theoretical_framework}. In the bottom of the field of view, a line of surface dimples ($\kappa>0$) coincides with a line of intense surface-attached vortices (large $|\omega_z|$) \new{which sit atop a downwelling ($\beta<0$). Nearby is a scar} – an elongated depression in the surface, likely related to a surface-parallel vortex \citep{aarnes_vortex_2025}: we can infer the opposite vertical velocity directions on either side of such vortex, where $\beta$ changes sign. Towards the centre of the image, two surface bulges ($\kappa<0$) coincide with two regions of upwelling ($\beta>0$). 

\new{For a given quantity $a(t,x,y)$ we compute the full space-time spectra $S_a^{(t,x,y)}(f,k_x,k_y)$ with a periodogram method, defined such that $\iiint S_a^{(t,x,y)} \mathrm{d} f \mathrm{d} k_x \mathrm{d} k_y = \overline{a^2}$. Here, $f$ is a temporal frequency and $k_x$ and $k_y$ are radian wavenumbers. We then integrate $S_a^{(t,x,y)}$ over the azimuthal angle to obtain $S_a^{(t,\lambda)}(f,k_\lambda)$, with $k_\lambda = \sqrt{k_x^2 + k_y^2} = 2 \pi / \lambda$, where $\lambda$ is a spatial scale. Finally, the 1-D spectra $S_a^{(\lambda)}$ and $S_a^{(t)}$ are obtained by integrating $S_a^{(t,\lambda)}$ over the temporal and spatial frequencies, respectively.}

Figures \ref{fig:spatial_spectra} (d-f) show the spatial spectra $S^{(\lambda)}_a$ of each quantity $a$, for five cases with increasing $\mathrm{Fr}$. The curvature spectra (figure \ref{fig:spatial_spectra} (d)) are amplified in magnitude as the turbulence increases in intensity, maintaining a similar shape with a peak around $1/\lambda \approx \SIrange{50}{60}{m^{-1}}$. Spectra of vertical vorticity (figure \ref{fig:spatial_spectra} (e)) are similar, with a slight shift towards higher wavenumbers as the turbulence is increased. The spectra of horizontal divergence in figures \ref{fig:spatial_spectra} (f) evidence greater energy at lower wavenumbers: compared to vorticity, the divergence field near the surface is characterized by somewhat larger features, as visible comparing figures \ref{fig:spatial_spectra} (b) and (c).

The spectra are normalized by the variances and the Taylor microscale in figure \ref{fig:spatial_spectra} (g-i), with peaks in the vorticity and surface curvature occurring at scales slightly larger than but comparable to $L_\mathrm{T}$. Probability density functions (PDFs) of each of the three quantities are shown in figure \ref{fig:spatial_spectra} (j-l). The curvature PDF is normalized in figure \ref{fig:spatial_spectra} (m) by the characteristic curvature scale $\overline{\beta^2}/g$ suggested by equation \ref{eq:kappa_g}, which provides a reasonable collapse of the data. In figure \ref{fig:spatial_spectra} (n-o), the distributions of the vertical vorticity and horizontal divergence are normalized by the Taylor timescale $T_\mathrm{T}$ (the timescale of an eddy of size $L_\mathrm{T}$ according to Kolmogorov scaling), which provides a measure of the fine timescales of the turbulence; the Kolmogorov time scale provides a similar collapse. The near-surface vertical vorticity has a greater magnitude than the horizontal divergence, in agreement with the surface measurements by \cite{qi_small-scale_2025}. There is a mild positive skewness to the horizontal divergence, consistent with the observation that upwellings tend to be more intense and organized than downwellings \citep{ruth_structure_2024}.

\begin{figure}
  \centerline{\begin{overpic}[width=1\linewidth]{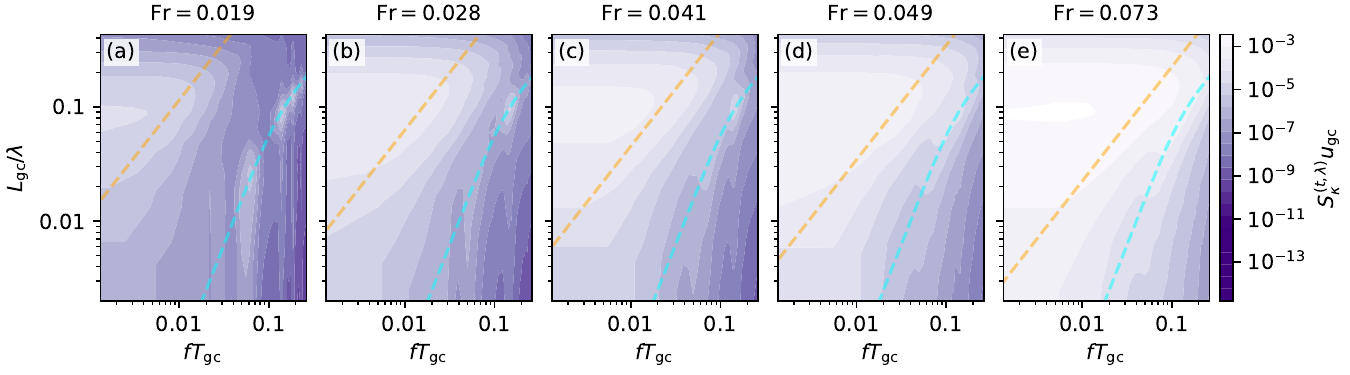}\end{overpic}}
  \caption{Space-time power spectral density of surface curvature for five cases spanning the range of Froude numbers. Dashed orange lines show features convected by the r.m.s.\ velocity; dashed blue lines show motions along the dispersion relation.}
\label{fig:measured_spacetime_spectra}
\end{figure}

In order to further explore the structure of the surface topography, in figure \ref{fig:measured_spacetime_spectra} we consider the space-time spectra of surface curvature, for the same sample cases considered in figure \ref{fig:spatial_spectra}. For \new{a consistent} normalization we use the gravity-capillary length $L_\mathrm{gc}$, time $T_\mathrm{gc} = (\sigma/(\rho g^3))^{-3/4}$ and velocity $u_\mathrm{gc}=L_\mathrm{gc}/T_\mathrm{gc}$. As the Froude number increases, so does the spectral density of the surface curvature. Most of the variance is concentrated in motions around, or slower than, the turbulent velocity $u'$, indicated by the dashed orange line. There is also some excess energy for motions around the capillary-gravity dispersion relation, shown as the dashed blue line. The generation and interaction of dispersive waves excited by subsurface turbulence is outside the scope of the present study and will be explored in the future.

Overall, in the present range of parameters, the magnitude and length scales of the surface curvature appear to respond to the features of the subsurface turbulence. In the following, the direct connection between surface deformation and subsurface flow is investigated in detail.

\subsection{Relationship between surface curvature and horizontal divergence}
\label{sec:variance_vs_variance}

\begin{figure}
  \centerline{\begin{overpic}[width=0.6\linewidth]{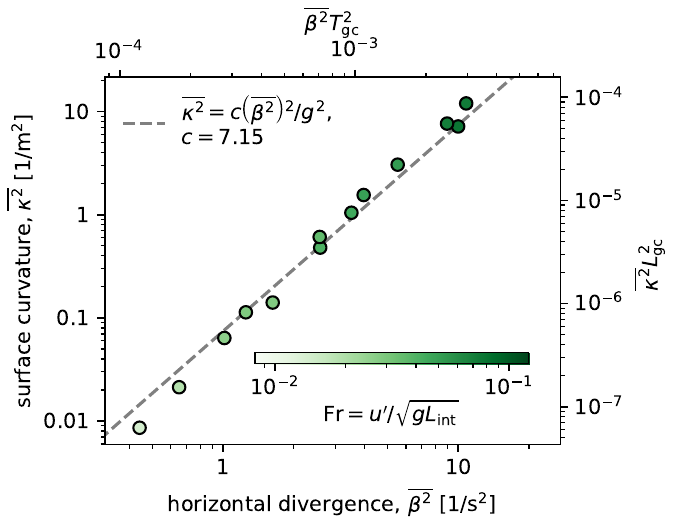}\end{overpic}}
  \caption{The scaling between the variances of the surface curvature and the horizontal divergence at $z=\SI{-5}{mm}$, exhibiting the power law predicted by equation \ref{eq:kappa_g}. Data are made dimensionless by the gravity-capillary length and time scales on the right and upper axes.}
\label{fig:variance_vs_variance}
\end{figure}

As anticipated in section \ref{sec:theoretical_framework}, from equation \ref{eq:kappa_g} we expect the surface curvature to scale with the square of the surface divergence when gravity is more important than surface tension. This is quantitatively confirmed in figure \ref{fig:variance_vs_variance}, where the r.m.s. curvature is shown to be proportional to the variance of the horizontal divergence measured at $z=\SI{-5}{mm}$ over a range of forcing conditions. Indeed, in the limit of zero surface tension, squaring both sides of (\ref{eq:kappa_g}) and averaging (denoted by an overbar), we obtain
\begin{equation}
    \overline{\kappa^2} \approx \frac{1}{g^2} \left( \overline{\left( \frac{D \beta}{D t} \right)^2} + \overline{\beta^4} + 4\overline{q^2} + \overline{\left[\textrm{cross terms} \right]} \right). \label{eq:kappa_variance_g}
\end{equation}
Given that the terms on the RHS are all small-scale quantities, we expect their ratios to be universal and therefore independent of the forcing. Thus, the entire RHS will be proportional to any individual term on the RHS, hence $\overline{\kappa^2} \propto \overline{\beta^4}/g^2$. As the kurtosis of the surface divergence is weakly dependent of the forcing \citep{qi_small-scale_2025}, we take $\overline{\beta^4} \propto (\overline{\beta^2})^2$ and arrive at $\overline{\kappa^2} \approx c \left(\overline{\beta^2}\right)^2 /g^2 $ (with $c$ a nondimensional prefactor), in excellent agreement with the data as shown in figure \ref{fig:variance_vs_variance}.

\begin{figure}
  \centerline{\begin{overpic}[width=1\linewidth]{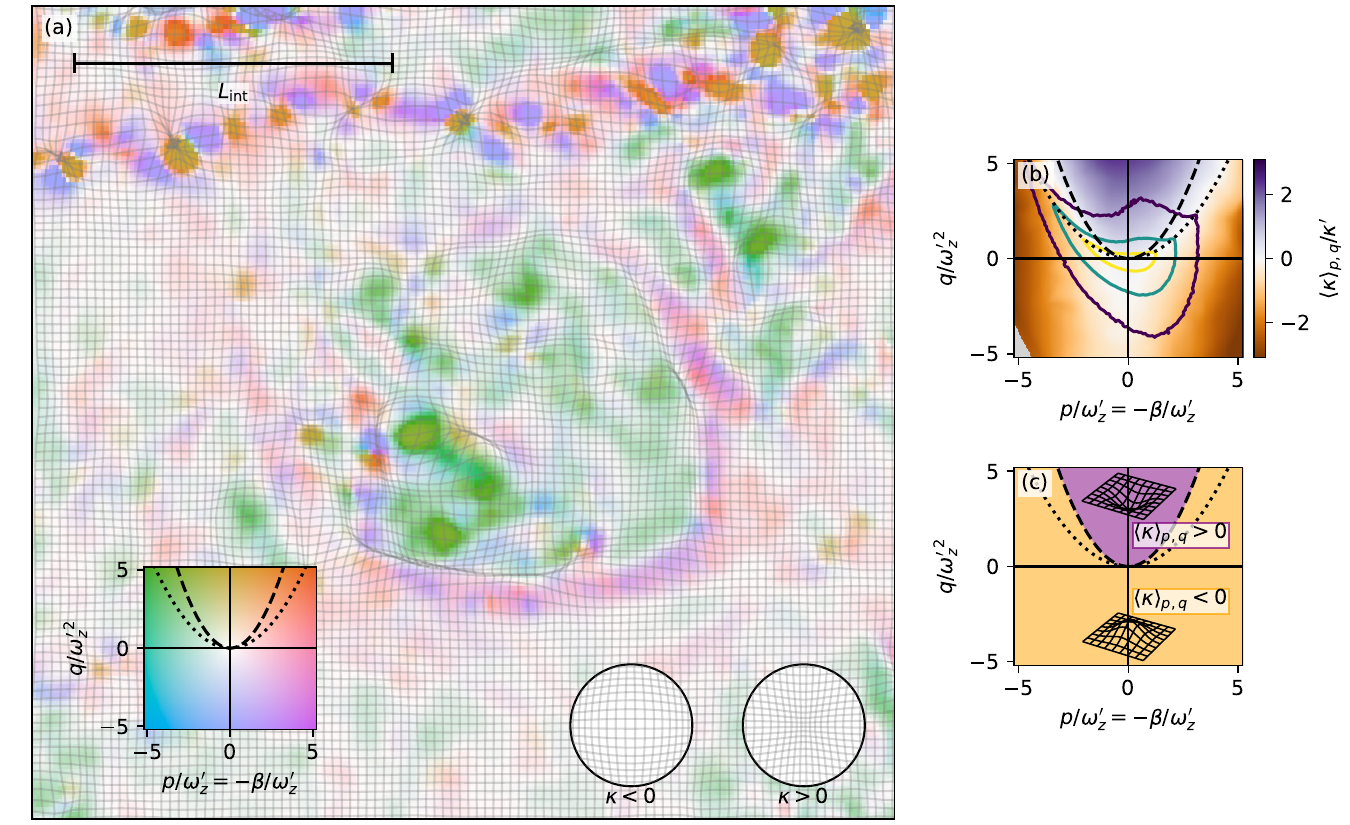}\end{overpic}}
  \caption{Relationship between the surface-parallel velocity gradient tensor invariants and the local surface morphology for the case with $\mathrm{Fr}=0.045$. (a) The invariants $p=-\beta$ and $q$ measured just beneath the surface, according to the color map in the bottom left. Deformed lines  give a BOS-inspired representation of the surface morphology: bulges ($\kappa<0$) and dimples ($\kappa>0$) are represented by lines spaced farther apart and closer together, respectively, as illustrated in the bottom right. Animated versions of this visualization for various $\mathrm{Fr}$ are included as a supplementary movie. (b) Contours of the JPDF of $p$ and $q$ at values 0.001, 0.01, and 0.1. The color in the background shows the mean value of the curvature conditioned on the combination of $(p,q)$ beneath. (c) A schematic representation of the surface topography in different portions of the $(p,q)$ map. The boundary $q=p^2/2$ derived from (\ref{eq:kappa_gs}), shown as the dashed line, separates regions where the curvature is typically positive and negative. The boundary $q=p^2/4$, shown as the dotted line, separates stable and unstable foci of the fluid motion \citep{perry_description_1987,cardesa_invariants_2013}.}
\label{fig:pq_deformations}
\end{figure}

Having shown a strong statistical relationship between the near-surface horizontal divergence and surface curvature, we now examine the local and instantaneous connection between fluid motions and the overhead curvature. Neglecting at first both surface tension and the unsteady \new{(in a Lagrangian sense)} term, from (\ref{eq:kappa_g}) we obtain
\begin{equation}
    \kappa_\mathrm{gs} = - \frac{\beta^2}{g} + \frac{2 q}{g} \label{eq:kappa_gs}
\end{equation}
with the subscript "gs" indicating “gravity-dominated and steady”. Under these assumptions, the surface curvature is related solely to the two invariants of the surface-parallel velocity gradient tensor. Their relation with the surface topography is illustrated in figure \ref{fig:pq_deformations}. In figure \ref{fig:pq_deformations} (a), a sample snapshot of the two-dimensional (2D) flow at $z=\SI{-2}{mm}$ and $\mathrm{Fr} = 0.045$ is depicted according to the colour scheme in the inset, associating different hues to different loci in the $p-q$ map. The instantaneous surface deformation is represented by grid lines deformed according to the refraction principle by which the BOS measurements operates. As sketched in the bottom right of the same panel, dilated lines represent bulges ($\kappa<0$), whereas lines coming together represent dimples $(\kappa>0$). Dimples tend to sit atop rotation-dominated regions (large positive $q$ with $p$ small in magnitude, coloured tan); bulges tend to sit atop regions of upwellings (large, negative $p$ values, green).

Statistical evidence of these associations is given in figure \ref{fig:pq_deformations} (b). Here, isolines of the $p-q$ JPDF are superposed to a colormap indicating the average value of $\kappa$ conditioned on the local value of the invariants beneath the surface. As schematically illustrated in figure \ref{fig:pq_deformations} (c), the region of positive curvature is bound by the condition $q=p^2/2$ (dashed line), in agreement with (\ref{eq:kappa_g}). This is distinct from the boundary $q=p^2/4$ (dotted line) which separates stable and unstable foci of the fluid motion \citep{perry_description_1987,cardesa_invariants_2013}.

\begin{figure}
  \centerline{\begin{overpic}[width=0.66\linewidth]{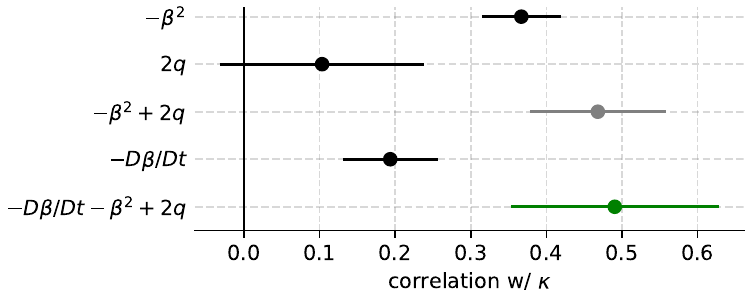}\end{overpic}}
  \caption{Correlations between terms related to the RHS of (\ref{eq:kappa_invariantdecomp}) and the measured value of $\kappa$. Circles and lines show the mean $\pm$ the standard deviation over all the forcing conditions for which we have fluid velocity just beneath the surface. Terms in black are those in the RHS of (\ref{eq:kappa_invariantdecomp}); the term in gray is the sum of the two steady terms (corresponding to $\kappa_\mathrm{gs}$ from (\ref{eq:kappa_gs})); the term in green is the entire RHS of EQ (1). The bar plots in the background give the p.d.f.s of the measured values.}
\label{fig:rhs_corr_w_kappa}
\end{figure}

We now consider how each term in the RHS of (\ref{eq:kappaz_transformed_invariantdecomp}) is correlated with the overhead surface curvature. Figure \ref{fig:rhs_corr_w_kappa} displays the respective Pearson correlation coefficients, \new{calculated with
\begin{equation}
    \mathrm{correlation}(\kappa,a) = \frac{\overline{(\kappa - \langle \kappa  \rangle_{t}) (a - \langle a  \rangle_{t}) }}{(\kappa - \langle \kappa  \rangle_{t})' (a - \langle a  \rangle_{t})'},
\end{equation}
using data in the middle of the domain where any parallax misalignment between the surface and sub-surface data is minimal \citep{weichert_thesis_2024}. Circles and error bars indicate the means and standard deviations, respectively,} among the nine cases for which we have flow measurements at $z=\SI{-2}{mm}$. The horizontal divergence $\beta$ exhibits the strongest correlation coefficient (about 0.37) with the surface curvature. This is further increased if $\beta$ is considered jointly with $q$: combining both terms leads to the modeled curvature $\kappa_\mathrm{gs}$ in (\ref{eq:kappa_gs}), whose correlation coefficient with the measured curvature is 0.47. Including also the unsteady term $-D \beta/D t$ leads to the modeled curvature $\kappa_g$ in (\ref{eq:kappa_g}), further improving the correlation coefficient between modeled and measured curvature to a value of 0.49. We note that the unsteady term is inherently noisy, as it involves temporal derivatives and second spatial derivatives, which are calculated numerically. Therefore, its contribution is possibly underestimated. 

The correlation between the measured and modelled surface curvature we obtain with this approach is one order of magnitude larger than those found by \cite{savelsberg_experiments_2009} in the flow behind an active grid, although the spatio-temporal scales and the depth of the velocity measurements in their study were comparable to ours. The difference is likely due to a combination of measurement accuracy (their surface scanning technique had a vertical resolution of \SI{0.1}{mm}, one order of magnitude coarser than BOS; \cite{moisy_synthetic_2009}), and the water surface being deformed by traveling waves generated at the grid.

\subsection{Modeling the gravity-capillary response of the surface }
\label{sec:modeling_gravcap}

\begin{figure}
  \centerline{\begin{overpic}[width=1\linewidth]{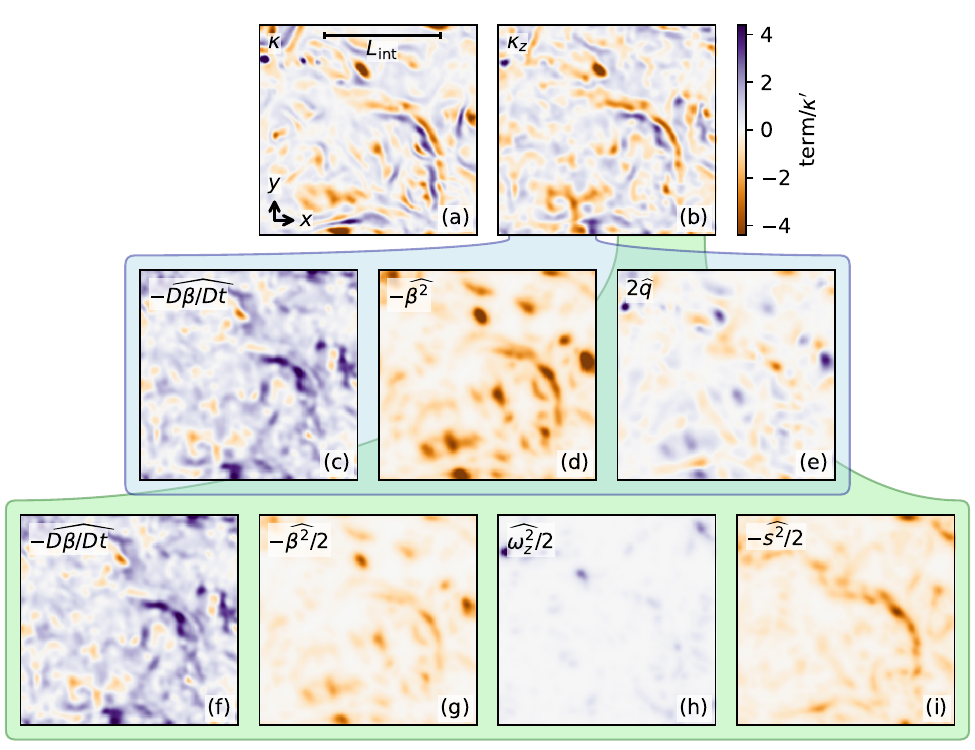}\end{overpic}}
  \caption{Modeling surface deformations based on the subsurface velocity field. The surface curvature with $\mathrm{Fr}=0.029$ as (a) measured and (b) modeled with velocity data just beneath the surface. (c-e) Three constituent terms which sum to $\kappa_z$  following (\ref{eq:kappaz_invariantdecomp}). (f-i) Four constituent terms which sum to $\kappa_z$  following (\ref{eq:kappaz}). An animated version of this figure is included as a supplementary movie.}
\label{fig:reconstruction_comparison_decomp}
\end{figure}

\new{Now,} we leverage the Euler-Laplace equation to directly model the surface curvature, accounting for both surface tension and gravity. Figure \ref{fig:reconstruction_comparison_decomp} compares, for the case $\mathrm{Fr} = 0.029$, a snapshot of the measured curvature $\kappa$ (a) to that of the modelled curvature $\kappa_z$ obtained using a spectral implementation of (\ref{eq:kappaz}) and velocity data from $z=\SI{-2}{mm}$ (b). The two agree remarkably well, especially at larger scales. Smaller scales are more heavily influenced by noise in the PIV data and, as will be discussed in greater detail below, by fine-scale flow features whose footprint does not reach the surface. Both displayed fields are normalized by the r.m.s.\ curvature at this condition, $\kappa'$, evidencing the close agreement in magnitude.

The modeled curvature $\kappa_z$ is the sum of three terms in (\ref{eq:kappaz_invariantdecomp}), shown in figures \ref{fig:reconstruction_comparison_decomp} (c-e). The Lagrangian rate of change of $\beta$ produces mostly regions of positive curvature \new{due to the slight mean flow in the facility}; stagnation due to upwelling and downwelling leads to negative curvature (bulges), and the competition between rotational and stretching motions embodied in $q$ yields both positive and negative curvature. For clarity, we also express $\kappa_z$ as the sum of the four terms in (\ref{eq:kappaz}), see figure \ref{fig:reconstruction_comparison_decomp} (f-i). Here we more directly see the effect of upwellings, downwellings and stretching motions leading to bulges, and vorticity leading to dimples. We follow this decomposition in the remainder of the paper.

\begin{figure}
  \centerline{\begin{overpic}[width=0.66\linewidth]{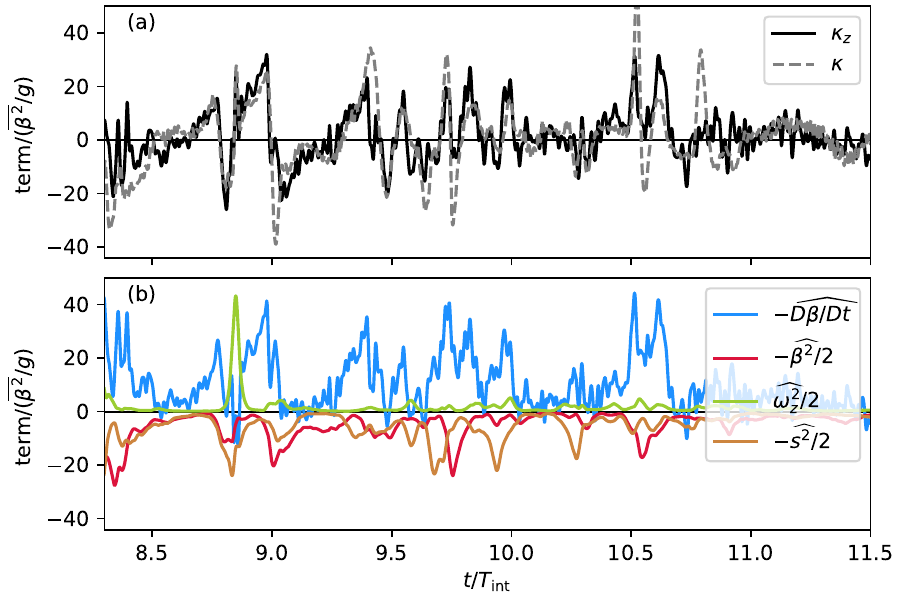}\end{overpic}}
  \caption{At one point in the middle of the domain for the case illustrated in figure \ref{fig:reconstruction_comparison_decomp}, timeseries of (a) measured and modeled curvature and (b) the terms constituting $\kappa_z$ through (\ref{eq:kappaz}). }
\label{fig:reconstruction_comparison_plotsvst}
\end{figure}

In figure \ref{fig:reconstruction_comparison_plotsvst} we show, for a point in the middle of the domain in the same case as figure \ref{fig:reconstruction_comparison_decomp}, a time series of measured and modelled curvature (a) and the contribution from the different terms (b). The time series corresponds to approximately 3 integral timescales of the turbulence, and the quantities are normalized by the characteristic curvature scale $\overline{\beta^2}/g$ suggested by (\ref{eq:kappa_g}). The divergence and stretching contribute to peaks of negative curvature, as discussed. The unsteady term exhibits the largest variance and, \new{as discussed above, a positive mean. Finally, enstrophy intermittently contributes to peaks in the surface curvature.}

\subsection{Scale dependence of the surface-subsurface relationship}
\label{sec:scale_dependence}

\begin{figure}
  \centerline{\begin{overpic}[width=1\linewidth]{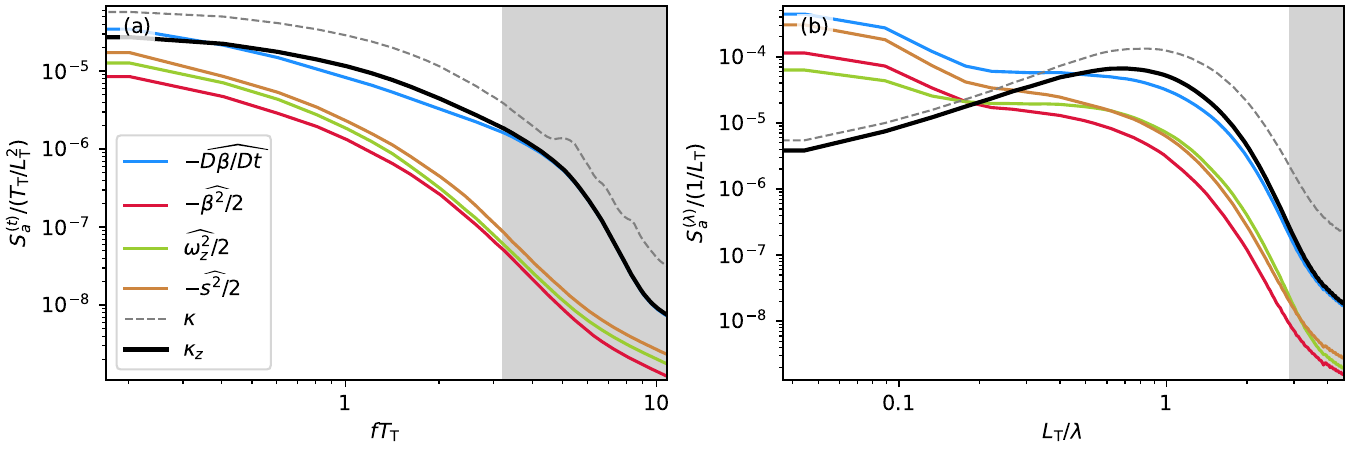}\end{overpic}}
  \caption{Power spectral density of the measured and modeled surface curvature and the constituent terms. (a) Temporal spectra. (b) Angular-integrated spatial spectra. Data is shown for $\mathrm{Fr}=0.029$, and the near-surface velocity yielding $\kappa_z$ is taken as close to the surface as possible, at $z=\SI{-2}{mm}$. Gray regions show where the velocity fields used to construct $\kappa_z$ are substantially filtered. }
\label{fig:measured_modeled_spectra_separate}
\end{figure}

Having considered the local and instantaneous link between surface curvature and the various terms contributing to it according to (\ref{eq:kappaz}), we now explore how this link is expressed at different spatial and temporal scales. In figure \ref{fig:measured_modeled_spectra_separate} we plot the temporal (a) and spatial (b) spectra of the measured and modelled curvature, as well as those of the different contributing terms, again for the sample case $\mathrm{Fr}=0.029$. Gray regions indicate spatial and temporal scales in the modelled data which are impacted by the Gaussian filter (specifically, the range where the filters exceed 50\% attenuation in spectral space). Measured and modelled curvature have very similar spatial and temporal spectra. Of the four terms contributing to $\kappa_z$, the unsteady term has the greatest variance, and its temporal spectrum is the closest to the one of the curvature. The spatial spectra show how all four terms possess high variance at large length scales. They are, however, significantly anti-correlated over such scales, resulting in a $\kappa_z$ spectrum peaked at relatively small scales and closely resembling the one of the measured curvature. This further highlights the ability of the present framework to capture not only the statistical correlation between surface deformation and specific observables in the subsurface flow, but also the spatio-temporal response of the surface topography to the subsurface topology.

\begin{figure}
  \centerline{\begin{overpic}[width=1\linewidth]{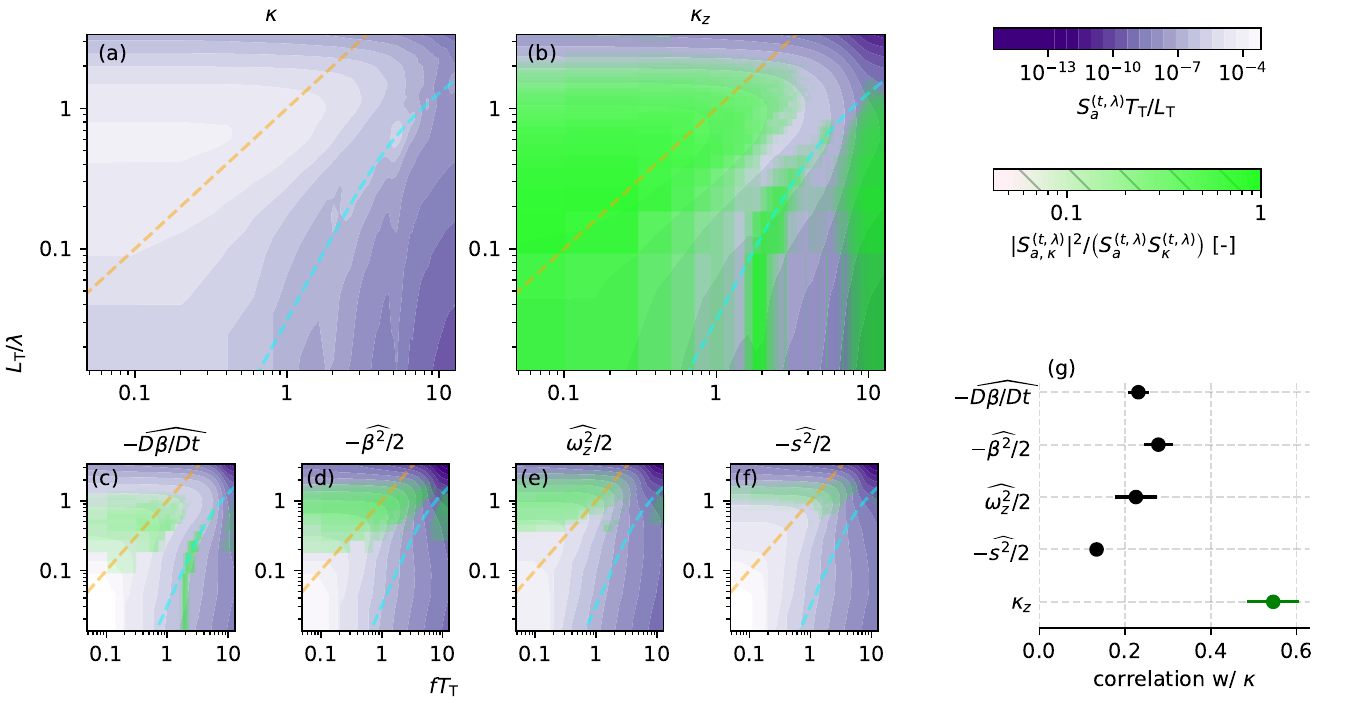}\end{overpic}}
  \caption{Spectral analysis of the measured and constructed curvature fields for the case with $\mathrm{Fr}=0.029$. Quantities are normalized by the Taylor length and time scales, $L_\mathrm{T}$ and $T_\mathrm{T}$. (a-b) Radially-integrated space-time power spectral density of the measured (a, $S_\kappa^{(t,\lambda)}$) and constructed (b, $S_{\kappa_z}^{(t,\lambda)}$) surface curvature fields. The semi-transparent green shading encodes the spectral coherence between $\kappa$ and $\kappa_z$. (c-f) As in (b), but for specific terms contributing to the modeled curvature field. (g) Among all the forcing conditions for which we have data at $z=\SI{-2}{mm}$ the mean $\pm$ the standard deviation of the correlations between $\kappa$ and either the modeled curvature $\kappa_z$ (in green) or individual terms in (\ref{eq:kappaz}) which sum to $\kappa_z$.}
\label{fig:measured_modeled_spectral_tr}
\end{figure}

The space-time spectral densities of the measured and modelled curvature are shown in figures \ref{fig:measured_modeled_spectral_tr} (a) and (b), respectively. Besides showing close resemblance, the figure also illustrates the extent of the spectral coherence between the two signals, defined as \new{
\begin{equation}
     \textrm{spectral coherence}(a,\kappa) = \frac{ \left| S_{a,\kappa}^{(t,\lambda)} \right|^2}{S_{a}^{(t,\lambda)} S_{\kappa}^{(t,\lambda)} }, \label{eq:spectral_coherence}
\end{equation}}
where $S_{a,\kappa}^{(t,\lambda)}$ is the cross-spectral density between $a$ and $\kappa$.

In particular, the signals are significantly correlated over spatio-temporal scales corresponding to the characteristic turbulent velocity $u'$ (dashed orange line) and propagating as gravity-capillary waves (dashed cyan line). The space-time spectra of the four contributing terms, and their spectral coherence with $\kappa$, are shown in figures \ref{fig:measured_modeled_spectral_tr} (c-f). The coherence between surface curvature and subsurface features is stronger at length scales $\mathcal{O}(10 L_\mathrm{gc}) \sim \mathcal{O}(L_\mathrm{T})$. The unsteady term is chiefly responsible for the spectral coherence between $\kappa_z$ and $\kappa$ along the dispersion relation. This is attributed to the orbital motions of traveling waves inducing alternating positive and negative patterns of horizontal divergence.

Finally, the Pearson correlations between the various terms and $\kappa$ are shown in figure \ref{fig:measured_modeled_spectral_tr} (g), again based on the nine cases for which PIV measurements at $z=\SI{-2}{mm}$ are available. As in figure \ref{fig:rhs_corr_w_kappa}, the horizontal divergence displays the higher correlation among the contributing terms. The modelled curvature $\kappa_z$ possesses a correlation coefficient of 0.54 with the measured curvature – a mild improvement over the value of 0.49 obtained for $\kappa_\mathrm{g}$, due to the inclusion of surface tension effects.

\subsection{Depth dependence of the surface-subsurface relationship}
\label{sec:depth_dependence}

\begin{figure}
  \centerline{\begin{overpic}[width=1\linewidth]{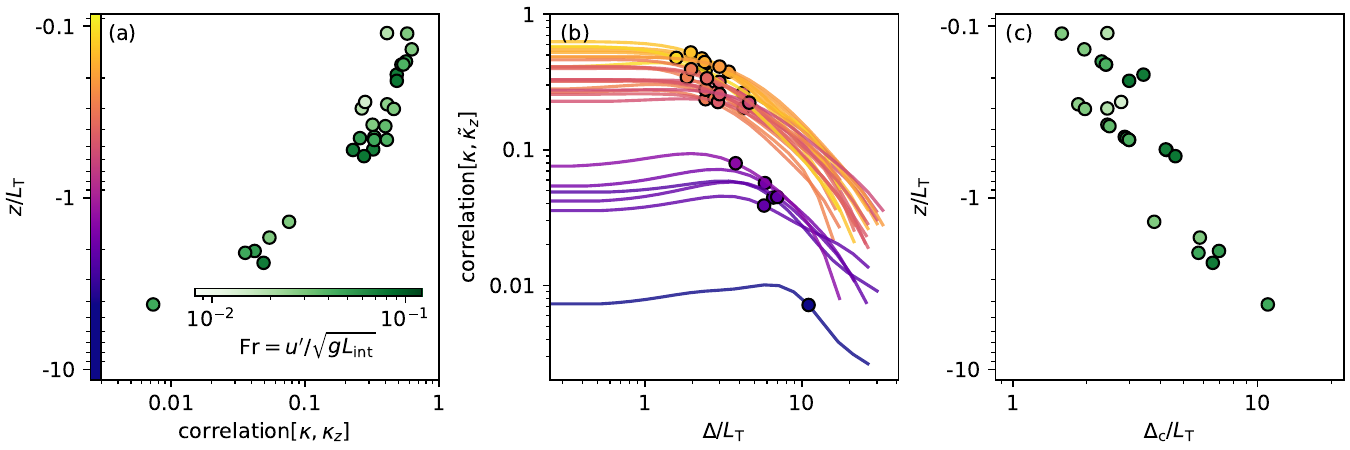}\end{overpic}}
  \caption{Correlation between $\kappa$ and $\kappa_z$. (a) The correlation as a function of depth normalized by the Taylor microscale. (b) Correlation at each depth, once the modeled curvature is low-pass filtered to a given scale. Circles denote $\Delta_\mathrm{c}$, the spatial scale at which the correlation begins to be hindered by the filter scale. (c) $\Delta_\mathrm{c}$ as a function of depth, suggesting that at deeper depths, a larger range of small scales in the velocity field are unrelated to the surface curvature. }
\label{fig:delta_c}
\end{figure}

So far, we have focused on the coupling between the surface curvature and the surface-parallel velocity gradients at a shallow depth of $z=\SI{-2}{mm}$. Because flow structures have a finite spatial extent and are not necessarily oriented vertically \citep{aarnes_vortex_2025}, one may intuit that the deeper a horizontal plane of fluid motion is beneath the surface, the less correlated it will be with the surface overhead. This expectation is confirmed by figure \ref{fig:delta_c} (a), which shows the correlation between the surface curvature $\kappa$ and the modelled curvature $\kappa_z$ as a function of $z$. The correlation decreases significantly with depth, dropping nearly an order of magnitude once $|z| \sim \mathcal{O}(L_\mathrm{T})$, \neww{which can be understood by considering the rapid decorrelation of $\beta(z)$ from the surface divergence $\beta_0$ within the upper-most Taylor microscale of the liquid \citep{babiker_experimental_2026}}.

To investigate the scale dependence of these correlations, in figure \ref{fig:delta_c} (b) we plot the correlation between the surface curvature $\kappa$ and the modelled curvature $\tilde{\kappa}_z$ low-pass-filtered with a Gaussian kernel of horizontal scale $\Delta$ \neww{(where the $\tilde{\cdot}$ operator denotes such filtering)}. The correlation remains mostly unaffected by filtering out the small scales up to $\Delta_\mathrm{c}$ (marked in the figure), which we conventionally define as the scale for which the rate of change of the correlation with respect to $\Delta$ attains the largest negative value. Evidently, $\Delta_\mathrm{c}$ increases with depth, as shown in figure \ref{fig:delta_c} (c): as the depth of the measured velocity field increases, so does the horizontal scale above which the modelled curvature resembles the measured surface curvature. Beyond some experimental scatter, the data indicates that information about the surface curvature at scales smaller than $\Delta$ can only be related to the flow structures within a depth $z \sim -b \Delta$, with $b \sim \mathcal{O}(1-10)$. This shall be taken as a mere order-of-magnitude estimate: more data over a broader range of conditions is needed to test more accurate and possibly nonlinear relations. 

\begin{figure}
  \centerline{\begin{overpic}[width=0.66\linewidth]{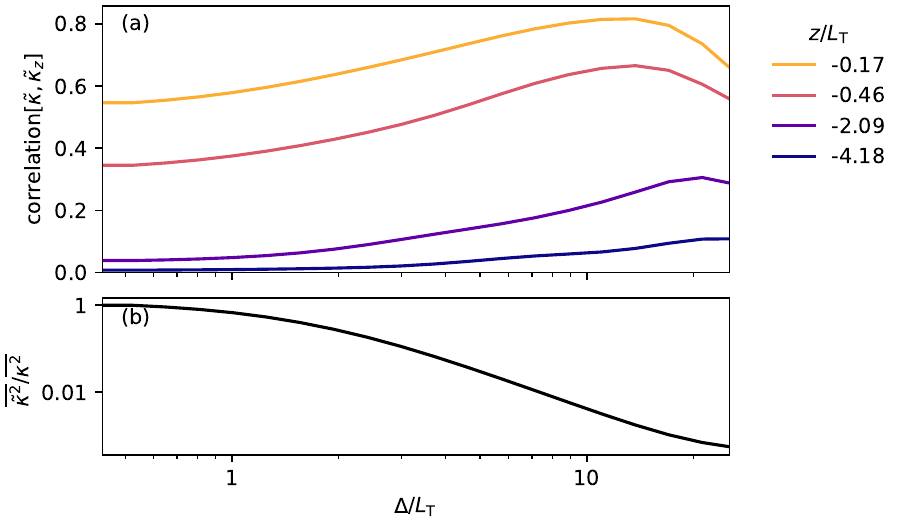}\end{overpic}}
  \caption{Impact of spatial filtering on the correlation between the measured and modeled curvature for $\mathrm{Fr}=0.045$. (a) At various depths, the correlation between $\kappa$ and $\kappa_z$ as a function of the spatial scale of the filter applied to both fields. (b) The decay of the variance of $\kappa$ with the low-pass filter scale $\Delta$. }
\label{fig:filtered_correlations}
\end{figure}

On the grounds of isotropy of the subsurface turbulence structures affecting the surface, we expect the velocity at greater depths to correlate with \new{only} increasingly large scales of the surface curvature field. This is confirmed in figure \ref{fig:filtered_correlations} (a) for the sample case $\mathrm{Fr}=0.045$, displaying the correlation between $\tilde{\kappa}_z$ and the measured curvature filtered at the same scale, $\tilde{\kappa}$, as a function of $\Delta$ and for different depths. \neww{At each depth, the sub-surface dynamics and surface curvature grow more highly correlated at larger spatial scales. This supports findings from \cite{babiker_experimental_2026}, who showed that when metrics for surface deformation and sub-surface divergence are spatially averaged over a horizontal domain spanning multiple $L_\mathrm{int}$, a strong instantaneous correlation between them persists through the entirety of the blockage layer. Quantities aggregated over smaller horizontal domains were seen to decorrelate more quickly with depth \citep{babiker_experimental_2026}.} While figure \ref{fig:filtered_correlations} (a) suggests an optimum filter size that grows with depth, one should consider the strong loss of information associated to spatial filtering. This is quantified in figure \ref{fig:filtered_correlations} (b), showing the variance of the filtered surface curvature $\tilde{\kappa}$ as a function of filter size: at $\Delta = 10 L_\mathrm{T}$, which yields maximal correlation at shallow depths, the \new{filtered} curvature variance is less than 1\% of the unfiltered measured level.

\section{Discussion and conclusions}
\label{sec:discussion_conclusions}

We have experimentally characterized how homogeneous turbulence deforms the free surface above it, both statistically and locally/instantaneously, across a range of Reynolds, Weber and Froude numbers. In particular, we have highlighted how the surface curvature naturally captures the topographic response of the surface to the underwater fluctuations. We have shown how the Euler-Laplace equation predicts the r.m.s.\ surface curvature to scale with the variance of the near-surface horizontal divergence measured at millimetric distance below the undisturbed surface. This prediction is found to be in excellent agreement with laboratory observations. As the near-surface divergence controls the fluid renewal from the bulk, this finding indicates that (at least in the present case without significant mean shear) the r.m.s.\ surface curvature is directly linked to the surface divergence, which in turn is directly correlated to the gas and heat transfer across the air-water interface. This may pave the way to inferring critical quantities such as the gas transfer velocity from surface elevation measurements. In this perspective, it is noteworthy that recent advances in field imaging are enabling unprecedented accuracy and resolution in measuring the ocean surface topography \citep{benetazzo_offshore_2012,laxague_e-pss_nodate}.

To investigate the local and instantaneous coupling between surface topography and subsurface flow structures, we have analysed the correlation of the surface curvature with the invariants of the velocity gradient tensor of the surface-parallel flow measured at various depths. When such depth is millimetric (i.e., smaller than the Taylor microscale), we find surface curvature to have a relatively high correlation coefficient with the (squared) divergence of 0.37. Physically, this is connected with bulges formed by the stagnation points atop upwellings and downwellings, which play major roles in free-surface turbulence \citep{guo_interaction_2010,ruth_structure_2024,wu_localised_2026}. The correlation increases to 0.54 when all terms contributing to the inviscid pressure disturbance according to the Euler-Laplace equation (i.e. vorticity, strain rate, rate-of-change of the horizontal divergence, and Laplace pressure) are used to model the surface curvature. Space-time spectra of the modelled and measured curvature demonstrate how the proposed framework allows \new{one} to reconstruct the local and instantaneous dynamics of the surface topography. 

\begin{figure}
  \centerline{\begin{overpic}[width=1\linewidth]{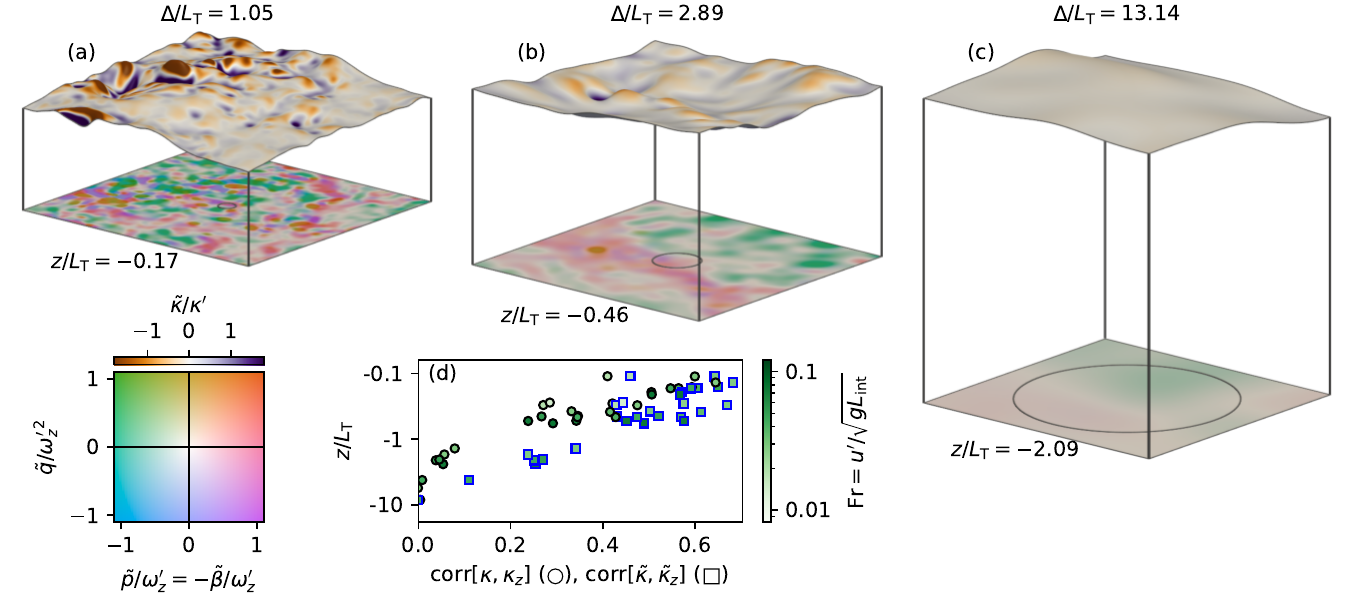}\end{overpic}}
  \caption{Comparison between the filtered surface curvature $\tilde{\kappa}$ and the filtered invariants of the subsurface velocity gradient tensor. (a-c) For various filter sizes $\Delta$, the filtered curvature field (top) as well as the subsurface velocity gradient invariants at a depth $z=-\Delta/(2\pi)$, themselves filtered to a scale $\Delta$. The physical sizes of the filters are represented by the diameters of the circles sketched in the planes. The depth of the subsurface planes are not shown to scale. (d) As a function of the depth beneath the surface, the correlation between the curvature and modeled curvature resulting from the flow at that depth. Circles show correlations between unfiltered curvature fields, and squares show the correlations between fields filtered to a scale $\Delta = 2 \pi |z|$.}
\label{fig:depthfilter_effect_3d}
\end{figure}

As mentioned, in the context of remote sensing, it would be attractive to invert the surface-subsurface relationship we have discussed: that is, leveraging measurements of the surface topography to infer the subsurface flow field. Based on the surface-subsurface decorrelation with depth, however, such an inference will likely only be possible down to spatial scales comparable to $|z|$. This is illustrated in figures \ref{fig:depthfilter_effect_3d} (a-c) in which surface curvature fields are low-pass filtered at increasingly large scales $\Delta$, displaying a dramatic decrease in signal variance. Beneath each curvature field is the map of the velocity gradient invariants at a depth $z=-\Delta/(2\pi)$, also smoothed to a scale $\Delta$. In figure \ref{fig:depthfilter_effect_3d} (d) we plot the correlation between $\kappa$ and $\kappa_z$ as a function of depth, and compare it to the one between the same fields low-pass-filtered at $\Delta = 2 \pi |z|$. Clearly, when we restrict ourselves to scales comparable to or larger than the depth of the velocity field, the correlation between modelled and measured surface curvature increases significantly.

Several extensions of this work are possible, for example including the effect of surfactants which strongly alter the surface and near-surface dynamics \citep{shen_effect_2004,yang_surfactant_2026}. Moreover, recent simulations have shed light on the geometry and dynamics of the intermittent layer above strong turbulence, whose surface does not only deform dramatically but can also break, generating bubbles and droplets \citep{calado_interfacial_2025,gaylo_quantifying_2026}. While these regimes clearly challenge the applicability of the present inviscid framework, they may offer other attractive avenues to infer the subsurface flow from the rich information associated with the properties of the dispersed phases.
\\
\\
\textbf{Funding.} Funding from the Swiss National Science Foundation (project \# 200021-207318) is gratefully
acknowledged.






\bibliographystyle{jfm}
\bibliography{references}


\end{document}